\documentclass[sigconf, nonacm]{acmart}
\AtBeginDocument{%
  }

\usepackage{multirow}
\usepackage{xcolor} 
\usepackage{threeparttable}
\usepackage{booktabs}

\usepackage{tabularx} 
\usepackage{pdflscape} 
\usepackage{pgfplots}

\usepackage[most]{tcolorbox}
\newcommand{\rqbox}[1]{
\begin{tcolorbox}[tile, size=fbox, boxsep=2mm, boxrule=0pt, top=0pt, bottom=0pt,
borderline west={1mm}{0pt}{blue!50!white}, colback=blue!5!white]
#1
\end{tcolorbox}
}

\newcommand{\tool}{Bamboo}

\begin{document}

\title{Bamboo: LLM-Driven Discovery of API-Permission Mappings in the Android Framework}

\author{Han Hu}
\affiliation{%
  \institution{Monash University}
  \city{Melbourne}
  \state{Victoria}
  \country{Australia}
}
\email{han.hu@monash.edu}

\author{Wei Minn}
\orcid{0000-0002-3191-9795}
\affiliation{%
  \institution{Singapore Management University}
  \city{Singapore}
  \country{Singapore}}
\email{wei.minn.2023@phdcs.smu.edu.sg}

\author{Yonghui Liu}
\orcid{0000-0001-7548-5100}
\affiliation{%
  \institution{Monash University}
  \city{Melbourne}
  \state{Victoria}
  \country{Australia}}
\email{Yonghui.Liu@monash.edu}

\author{Jiakun Liu}
\orcid{0000-0002-7273-6709}
\affiliation{%
  \institution{Singapore Management University}
  \city{Singapore}
  \country{Singapore}}
  \email{jkliu@smu.edu.sg}

\author{Ferdian Thung}
\orcid{0000-0002-5566-3819}
\affiliation{%
  \institution{Singapore Management University}
  \city{Singapore}
  \country{Singapore}}
  \email{ferdianthung@smu.edu.sg}

\author{Terry Yue Zhuo}
\orcid{0000-0002-5760-5188}
\affiliation{%
  \institution{Monash University}
  \city{Melbourne}
  \state{Victoria}
  \country{Australia}}
\email{terry.zhuo@monash.edu}

\author{Lwin Khin Shar}
\orcid{0000-0001-5130-0407}
\affiliation{%
  \institution{Singapore Management University}
  \city{Singapore}
  \country{Singapore}}
  \email{lkshar@smu.edu.sg}

\author{Debin Gao}
\orcid{0000-0001-5130-0407}
\affiliation{%
  \institution{Singapore Management University}
  \city{Singapore}
  \country{Singapore}}
  \email{lkshar@smu.edu.sg}

\author{David Lo}
\orcid{0000-0002-4367-7201}
\affiliation{%
  \institution{Singapore Management University}
  \city{Singapore}
  \country{Singapore}}
  \email{davidlo@smu.edu.sg}

\renewcommand{\shortauthors}{Han et al.}
\newcommand{\han}[1]{\textcolor{red}{Han Hu: {#1}}}

\begin{abstract}
The permission mechanism in the Android Framework is integral to safeguarding the privacy of users by managing users' and processes' access to sensitive resources and operations. 
As such, developers need to be equipped with an in-depth understanding of API permissions to build safe, robust and functional Android applications (apps). 
Unfortunately, the official API documentation by Android chronically suffers from imprecision and incompleteness, causing developers to spend significant effort to accurately discern necessary permissions. This potentially leads to incorrect permission declarations in Android app development, potentially resulting in security violations and app failures.
Recent efforts in improving permission specification primarily leverage static and dynamic code analyses to uncover API-permission mappings within the Android framework. 
Yet, these methodologies encounter substantial shortcomings, including poor adaptability to Android Software Development Kit (SDK) and Framework updates, restricted code coverage, and a propensity to overlook essential API-permission mappings in intricate codebases.
This paper introduces a pioneering approach utilizing large language models (LLMs) for a systematic examination of API-permission mappings, scanning all Java methods within the Android SDK to ascertain required permissions, significantly enhancing traditional methods in terms of code coverage, accuracy, and adaptability. In addition to employing LLMs, we integrate a dual-role prompting strategy and an API-driven code generation approach into our mapping discovery pipeline, resulting in the development of the corresponding tool, \tool{}. We formulate three research questions to evaluate the efficacy of \tool{} against state-of-the-art baselines, assess the completeness of official SDK documentation, and analyze the evolution of permission-required APIs across different SDK releases. Our experimental results reveal that \tool{} identifies 2,234, 3,552, and 4,576 API-permission mappings in Android versions 6, 7, and 10 respectively, substantially outperforming existing baselines, Dynamo and Arcade, by 86.48\%, 100\%, and 77.85\%. 
Additionally, it uncovers over 3,000 significant permission declaration omissions in the official documentation across Android 7, 10, and 15, highlighting considerable gaps in its completeness. 
\end{abstract}



\received{20 February 2007}
\received[revised]{12 March 2009}
\received[accepted]{5 June 2009}

\maketitle

\section{Introduction}
The Android application (app) framework is integral to the security and integrity of millions of mobile devices globally. 
Permissions in the Android Framework play a vital role in the preserving security of millions of mobile devices running Android globally.
Specifically, the Android Framework uses a permission system to manage users and processes' access to sensitive data and operations inside the device. 
For developers, understanding the specific permissions required by Android Application Programming Interface (APIs) is important when developing secure apps.
However, developers have to contend with the official Android documentation which has a reputation for its inconsistency and also incompleteness~\cite{wang2022runtime, pscout, android_reference}. 
This lack of clarity can lead to errors in permission declarations, potentially causing app failures and compromising user privacy~\cite{luo2013real}.


For instance, consider a social media app equipped with features that enable users to share multimedia content such as videos and audio immediately after capturing them with smartphone cameras. 
Due to inaccuracies or omissions in the Android Software Development Kit (SDK) documentation, there is often confusion among app developers regarding the necessary permissions for camera access, multimedia processing, and recording functionalities. 
Specifically, if they assume that the \texttt{CAMERA} permission implicitly includes the \texttt{RECORD\_AUDIO} permission when recording videos, it could lead to app crashes. 
When an app lacks the required permission and still invokes the API, the system throws an unhandled security exception to prevent unauthorized access, causing the app to terminate unexpectedly.

More importantly, a lack of clear understanding among developers about the exact permissions required for the APIs they utilize can lead to the declaration of excessive or irrelevant permissions. 
These practices have been shown to be in violation of the Principle of Least Privilege and have been demonstrated to lead to more severe consequences: they not only inflate the size of the app but also introduce significant security vulnerabilities~\cite{bugiel2012towards, CloakandDagger}. 
Such imprecise granting of permissions leads to increase in potential attack vectors where malicious apps can be designed to exploit these superfluous permissions to perform various kinds of attacks such as privilege escalation and component hijacking. 
Thus, this example motivates the critical need for precise and comprehensive permission-API mappings for Android SDK.
Accurate permission-API mappings are crucial among many tasks. 
For instance, investigations into Security Policy Compliance~\cite{pscout}, Permission Misuse Detection~\cite{bartel2012automatically}, and Refinement of Permission Granularity~\cite{felt2011effectiveness} critically depend on precise mappings to effectively validate their results. Areas such as Static Analysis for Security Auditing~\cite{octeau2013effective} and Behavioral Analysis for Context-Aware Permissions~\cite{roesner2012user} rely on these accurate mappings to maintain the integrity and relevance of their conclusions. The absence of accurate mappings could lead to incorrect security assessments and flawed app permissions, threatening both user privacy and system integrity.


Existing works in the topic of API-permission mapping has primarily relied on static code analysis of the Android Framework~\cite{pscout, axplorer, arcade, psgen, natidroid},  and dynamic analysis ~\cite{stowaway, dynamo}.
Although our understanding of the Android permission mechanism and API-permission mapping has greatly improved, the limitations of static~\cite{samhi2024call, 10.1145/3647632.3651388, li2017static} and dynamic analysis~\cite{wang2018empirical,akinotcho2024mobile} still affect the accuracy and completeness of these mappings.
For example, static analysis may fail to capture runtime permissions dynamically assigned based on user interactions or system conditions, whereas dynamic analysis often suffers from limited code coverage, missing out on rare or context-specific execution paths.
Furthermore, both methods struggle to adapt to the frequent changes and expansions in the Android SDK in terms of compatibility issues,  often leading to outdated or incomplete analyses~\cite{dynamo}. As a result, many API-permission mappings remain unmaintained and thus inaccurate for newer Android releases with changes in security policies.

To address existing challenges, we propose a novel three-phase pipeline that integrates large language models (LLMs) into the discovery of API-permission mappings within the Android Framework.
In the first phase, we extract all Java APIs across the SDK. In the second phase, we analyze these extracted APIs using LLMs through our proposed dual-role prompting strategy. Finally, in the third phase, we employ API-driven LLM code generation to produce self-contained test cases for selected APIs, thereby verifying the detected permission-required APIs.
Our approach leverages the advanced code comprehension and code generation capabilities of LLMs, enabling a more general, thorough, and up-to-date analysis of API-permission relationships compared to traditional methods. Unlike existing approaches, our method systematically analyzes all Java methods across the SDK using LLMs, significantly improving the code coverage in SDK of permission mapping analysis. We implement this pipeline and develop a corresponding tool, \tool{}, by integrating the complete three-phase pipeline.
To evaluate the efficacy of \tool{} across various scenarios, we define the following research questions:

\begin{itemize}
    \item \textbf{RQ1}: How effective is \tool{} compared to existing works?
    \item \textbf{RQ2}: How well does \tool{} perform when evaluated against Android SDK Source Code Annotation and Documentation?
    \item \textbf{RQ3}: What insights can \tool{} provide about API-permission mappings across major Android Framework releases?
\end{itemize}

In \textbf{RQ1}, we conduct comparative experiments with established baselines Dynamo~\cite{dynamo} and Arcade~\cite{arcade}. The experimental results indicate that \tool{} identifies 2,234, 3,552, and 4,576 API-permission mappings in Android versions 6, 7, and 10, respectively, substantially outperforming existing baselines by 86.48\%, 100\%, and 77.85\%. These results strongly demonstrate the effectiveness of our tool.
In \textbf{RQ2}, we evaluate Android's developer documentation from \tool{}, identifying significant gaps and inaccuracies that shed light to the current state of API documentation. \tool{} discovers 3,487, 3,906, and 2,202 unannotated permission-required APIs in the source code of Android versions 7, 10, and 15. Additionally, \tool{} identifies 3,539, 4,519, and 3,100 non-standardized permission-required APIs~\footnote{'Non-standardized' means that the API does not declare the required permission according to the Android documentation specifications. We will explain details in RQ2.}  in the official Android online documentation for versions 7, 10, and 15. These findings highlight considerable security risks in the existing Android official documentation.
Finally, in \textbf{RQ3}, by observing and analyzing discrepancies and implications in API-permission mappings across several major SDK updates, we summarize potential reasons and uncover evolving trends and previously undetected features within SDK development. This analysis generates crucial insights that significantly influence the future usage of these SDK APIs.

Overall, this paper introduces a robust and adaptable API-permission mapping tool \tool{}, which pushes the literature of Android permission analysis in terms of precision and completeness of the API-permission mapping, and insights into the existing mappings and the official mapping documentation. 
The contributions of this paper are threefold:
\begin{enumerate}
    \item To our knowledge, this is the first study integrating LLMs into the analysis of Android API permissions, developing a tool \tool{} that identifies a broader array of API-permission mappings than existing baselines. Our code is open-sourced to enhance the research community~\footnote{https://github.com/huhanGitHub/LLMPerm}.
    \item Our methodology surpasses traditional static and dynamic approaches in terms of flexibility and effectiveness, working well across multiple SDK versions and Android runtime environments.
    \item We conduct an empirical study that not only evaluates the current state of API-permission mappings but also uncovers the inaccuracies and incompleteness of current Android official documents, providing deeper insights into this critical field and revealing underlying patterns and potential vulnerabilities.
\end{enumerate}

We organize this paper as follows: we first introduce related works in Section~\ref{sec:relatedWork}. Second, we present our LLM-Driven API-Permission Mapping Discovery pipeline in Section~\ref{sec:method}.
Third, we investigate three RQs in Section~\ref{sec:evaluation}. Finally, we discuss threats to validity of our approach and experiments in Section~\ref{sec:discussion}.

\section{Related Work}
\label{sec:relatedWork}

\subsection{Android API-Permission Mapping}

\subsubsection{Dynamic Analysis-based Analysis}
The first work to study Android Permission is Stowaway~\cite{stowaway}. It leverages feedback-directed fuzzing, an dynamic analysis approach, to invoke API calls that an app uses, and maps those API calls to permissions. 
HeapHelper~\cite{heaphelper} performs heap memory snapshot analysis that leverages the dynamic information stored in the heap of Android Framework execution to assist in generating a more precise call graph that model the runtime behavior of procedures inside the Android Framework. Precise call graphs allow for fewer false positives permissions being mapped to the framework APIs which leads to a more usable mapping for security analysis.
Dynamo~\cite{dynamo} revisits the app dynamic analysis technique and the imprecision issue in existing static analysis approaches for Android API-permission mapping, and delivers further improvement in both precision and code coverage upon the existing works. It achieves better coverage by employing static analysis to form semantically relevant seed input values based on the parameters' names in the API signature, and testing strategies that include returned security check message in its feedback so that it can bypass failing security checks for further explorations down the same execution path. It also delivers robustness by dynamically instrumenting memory to obtain the state of the procedure-under-test, compared to Stowaway which requires the modification of the Android Open Source Project (AOSP) code in order to hook the permission checking mechanism of Android Framework.
Given the open-source nature and demonstrated effectiveness of the tool, we selected Dynamo, the latest state-of-the-art approach, as one of the baselines for our experiment.

However, since dynamic analysis requires real-time execution, Android API-permission mappings built using this approach depend heavily on specific, rare, and untested execution paths. As a result, permission checks triggered in these scenarios are often missed, leading to incomplete mappings.
In addition to low coverage, dynamic analysis techniques also suffer from other shortcomings such as inefficiency (slow convergence), and lack of robustness (needing extensive setup for newer environments and releases). 
To address the issue of low coverage, \tool{} analyzes the entire Android SDK source code statically to derive extensive execution scenarios without the need to wait for execution to dynamically reach it. 
On top of improving efficiency, advanced language modeling enables more accurate and in-depth prediction and interpretation of permission scenarios, providing a comprehensive and resource-efficient solution for permission analysis.


\subsubsection{Static Analysis-based Approaches}
PScout~\cite{pscout} is the first work to extract the permission specification from the Android OS source code using static analysis. Its aim is to provide better coverage in contrast to Stowaway~\cite{stowaway}, the sole existing approach at that time, which is based on dynamic analysis and consequently suffers from low code coverage. However, PScout suffers from imprecision issues that is common in many other static analysis approaches. 
Axplorer~\cite{axplorer} attempts to improve precision by conducting a systematic study on the design pattern peculiarities of Android Framework code such as message-based IPC, and framework component interconnection. 
Arcade~\cite{arcade} adds path-sensitivity to static analysis for further precision improvement based on a novel graph abstraction technique. Arcade extracts Control Flow Graph representation of Android Framework, and derives a novel Access-Control Flow Graph which is processed to produce a succint representation of the access control conditions enforced by the API in the form of first-order logic.
PSGen~\cite{psgen} extends permission mapping analysis to native framework APIs in Android NDK in contrast to existing works that only performs permission specification for Java Framework APIs only. 
Natidroid~\cite{natidroid} performs permission analysis in cross-language scenarios i.e. between Framework API in Java, and permission check inside native code. It identifies Android Interface Definition Language and Java Native Interface patterns, two major Java-native communication patterns, inside both Java and native code of Android Framework to extract entry points and construct a comprehensive Interprocedural Control Flow Graph (ICFG) for a more complete permission specification analysis.
Given the open-source nature and demonstrated effectiveness of the tool, we selected Arcade, the latest state-of-the-art tool, as one of the baselines for our experiment.

While static analysis techniques for Android API-permissions mappings scan the Android Framework's codebase and extract permissions required from call graphs and control-flow graphs without executing anything, they miss context-dependent (message handlers that triage message depending on message code) permissions leading to false positives that mistakenly lables APIs to require permissions that they actually do not require.
As the static analysis tools have to be manually implemented by the researchers, they may misinterpret complex code structures or even fail to consider specific design patterns, leading more incomplete mappings compared to those of dynamic analysis. 
This also means that any updates to Android Framework would require a revamp in the implementation of the static analysis tools to accommodate those changes to maintain accuracy in building runtime models of the framework.
\tool{} enhances static analysis by leveraging LLM that has the capacity for a deeper and more accurate interpretation of the Android Framework source code. 
Unlike traditional static analysis, our method is robust against frequent SDK and Framework updates, as it is not reliant on specific code structures or design patterns. 
This allows for continuous and reliable permission analysis without the need for frequent adjustments, and providing the Android app developers and security analysts with precise, comprehensive and up-to-date API-permission mappings that is crucial for thorough code inspections and malware detection.

\subsection{LLM for SE}

\subsubsection{Software Testing with LLM}

A survey by Wang et al.~\cite{llmsoftwaretestinglandscape} taxonomizes works that applies LLMs in the topic of Software testing into unit test case generation, test oracle generation, and system test input generation. Our approach is in line with works under the system test input generation category~\cite{ye2021automated, taesiri2022large, shrestha2021slgpt, hu2023augmenting, hu2024enhancing, hu2023automated,
mathur2023automated, zimmermann2023automating, taeb2024axnav, luu2023can,
khanfir2023efficient, deng2023large, deng2023zeroshot, ackerman2023fuzzingparsers, 
yu2023llmtestscriptgeneration, Deng2023PentestGPTAL, 
sun2023smtsolvervalidation, deng2023targetautomatedscenariogeneration, Zhang_2024, 10.1145/3597503.3639121, 
tsigkanos2023variablediscovery, yang2024whitefox, 10.1145/3708528} as our approach is not concerned about functional verification of software and thus is not related to oracles, and the test case we generate are not concerned with individual procedures inside the Android Framework but rather high-level framework APIs that abstracts away the many unit-level procedures. The existing works are concerned with generating test cases for Android apps, deep learning library, compilers, SMT solvers, cyber-physical systems, and so on. To our best knowledge, our work is the first to approach Android Framework API-permission mapping problem by leveraging LLM when generating test cases for framework APIs.

\subsubsection{Code Generation with LLM}
Code generation involves the automated creation of executable code from software requirements \cite{liu2022deep}. 
Traditionally, code generation relies on predefined rules, templates, or configuration data and, hence, have faced significant limitations when it comes to flexibility \cite{halbwachs1991generating, icse_Whalen00}. 
The emergence of deep learning and LLMs has revolutionalizaed the landscape of code generation. Existing extensive code corpora has enabled recent works to focus on training LLMs that are designed for more complex code generation challenges \cite{liu2022deep, acl_ZanCZLWGWL23}.
Several LLMs such as Codex \cite{chen2021evaluate}, CodeGen \cite{iclr_NijkampPHTWZSX23}, StarCoder \cite{lozhkov2024starcoder}, CodeLlama \cite{Baptiste2023Codellama}, and DeepSeek-Coder \cite{guo2024deepseek} have demonstrated exceptional capabilities in terms of efficiency and accuracy of  synthesizing executable code.
In our work, we leverage two cutting-edge techniques, In-Context Learning~\cite{rubin2021learning} and Multi-Role Player prompting~\cite{dong2022survey}, as integral strategies in our LLM-driven code generation and analysis pipeline. 





\section{Methodology}
\label{sec:method}

\begin{figure}[htbp]
\centering
\includegraphics[width=\linewidth]{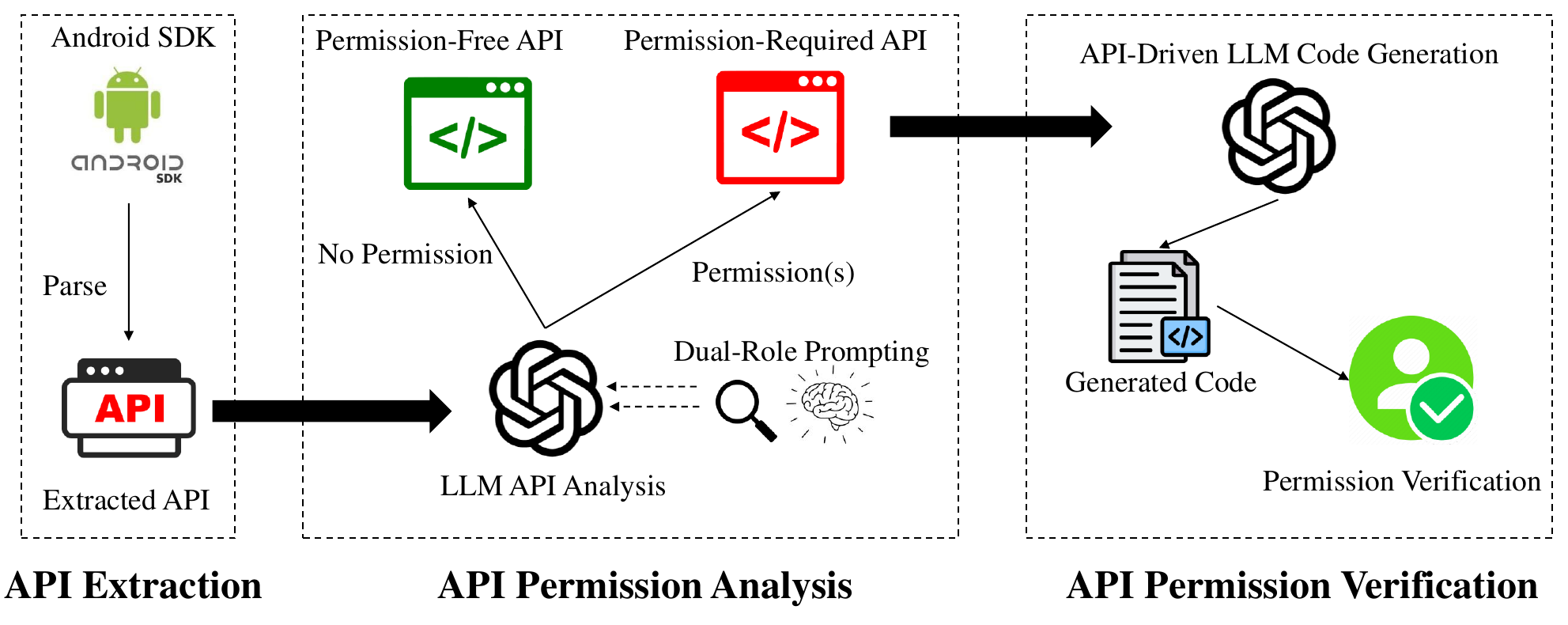}
\caption{LLM-Driven API-Permission Mapping Discovery Pipeline}
\label{fig:pipeline}
\end{figure}

Figure~\ref{fig:pipeline} depicts the overview of the \tool{}'s pipeline which is organized into three primary phases: Android SDK API Extraction, LLM-based API Permission Analysis, and API Permission Verification.

In the first phase, Android SDK API Extraction, \tool{} identifies the full set of the Android SDK's APIs, for our analysis to cover far corners of the Android Framework that are less documented or rarely used. 
We extract all applicable APIs from the Android SDK source code through the use of static analysis techniques such as AST parsing and keyword matching. 

In the second phase (LLM-based API Permission Analysis), each extracted API undergoes a rigorous examination conducted by a specially tailored LLM. 
The LLM leverages custom-designed dual-role prompting strategy to analyze and interpret both the API code and the comments. 
This step enables the LLM to accurately determine whether each API requires specific Android permissions, thereby addressing the complexities associated with API permission specifications.


Finally, the API Permission Verification phase validates the permission predictions by applying our API-driven LLM code generation technique. This technique leverages an LLM to automatically generate initial self-contained test cases for APIs that have been initially identified as requiring permissions.
Manual guidance are still involved for occasional human-guided refinement of the generated test cases to ensure that they are both accurate and suitable for the specific permission requirements and scenarios under evaluation. 
This is due to the inherent variability and instability of LLM outputs.
The verification of the generated test cases is done by executing them within demo apps that act as the API clients of the Android Framework on an emulated Android phone. 

\subsection{Android API Extraction} 
Extraction of APIs from the Android SDK is a crucial step to ensure the extensive coverage of our permission analysis.
This phase has two steps: 1) method signature extraction and 2) contextual information extraction. Together, these steps pre-process dataset for subsequent phases of \tool{}.

\subsubsection{Step 1: Method Signature Extraction} 
The APIs of the Android SDK are identified by unique method signatures that are invoked by client Android applications. 
As such, to extract all the relevant APIs within the Android SDK, we employ a combination of abstract syntax tree (AST) parsing and keyword matching techniques.

\paragraph{Parsing AST} 
We construct a tree representation of Java code elements inside the Android SDK by analyzing the AST of the source code which enables us to examine each node within the tree that corresponds to elements such as classes, methods, and control statements.
This systematic traversal of the ASTs ensures an accurate identification of all method signatures that comprise their respective parameter lists and scopes. 
In our implementation, we employ the \texttt{javalang} Python library as the AST parser.

\paragraph{Keyword Matching} 
As the AST parser may raise an exception upon encountering improperly formatted or incomplete methods within the SDK source code, we complement AST parsing with a keyword matching approach tailored for Java syntax and identifiers. 
This method identifies method declarations in terms of 1) access modifiers (\texttt{public}, \texttt{protected}, and \texttt{private}), 2) return types (\texttt{void}, \texttt{int}, \texttt{String}, and other common data types), and 3) annotations (\texttt{@Override}, \texttt{@Deprecated}, and \texttt{@RequiresPermission}).
The matching is also done for specific keywords inside the method name that suggest related permissions or certain interactions with system features (\texttt{get}, \texttt{set}, \texttt{create}, \texttt{request}, and \texttt{manage}). 
These keywords are derived based on Java syntax rules, empirical naming conventions observed in the Android SDK source code, and the authors' domain knowledge.
Combining keyword matching with AST parsing enables effective detection by capturing both documented and undocumented methods within the Android SDK, which might otherwise be overlooked by approaches that do not analyze the SDK directly.

\subsubsection{Step 2: Contextual Information Extraction} 
The extracted method signature needs to be complemented by additional contextual information related to the identified APIs. 
This information includes the API level (version of the Android Framework/SDK), deprecation status, and any other special descriptions that accompany the method inside the SDK source code. 
Contextual information helps align the usage of the APIs with specific Android versions in an effort to enhance the robustness of permission analysis for future Android Frameworks.

Identified APIs are organized into a structured database, with each entry documenting the method’s signature, its location within the SDK, and any associated permissions if applicable.
The database is part of the engineering effort to allow for the automation of LLM-based API Permission Analysis conducted in the subsequent phase.

\subsection{LLM-based API Permission Analysis}
\begin{figure}[htbp]
\centering
\includegraphics[width=0.9\linewidth]{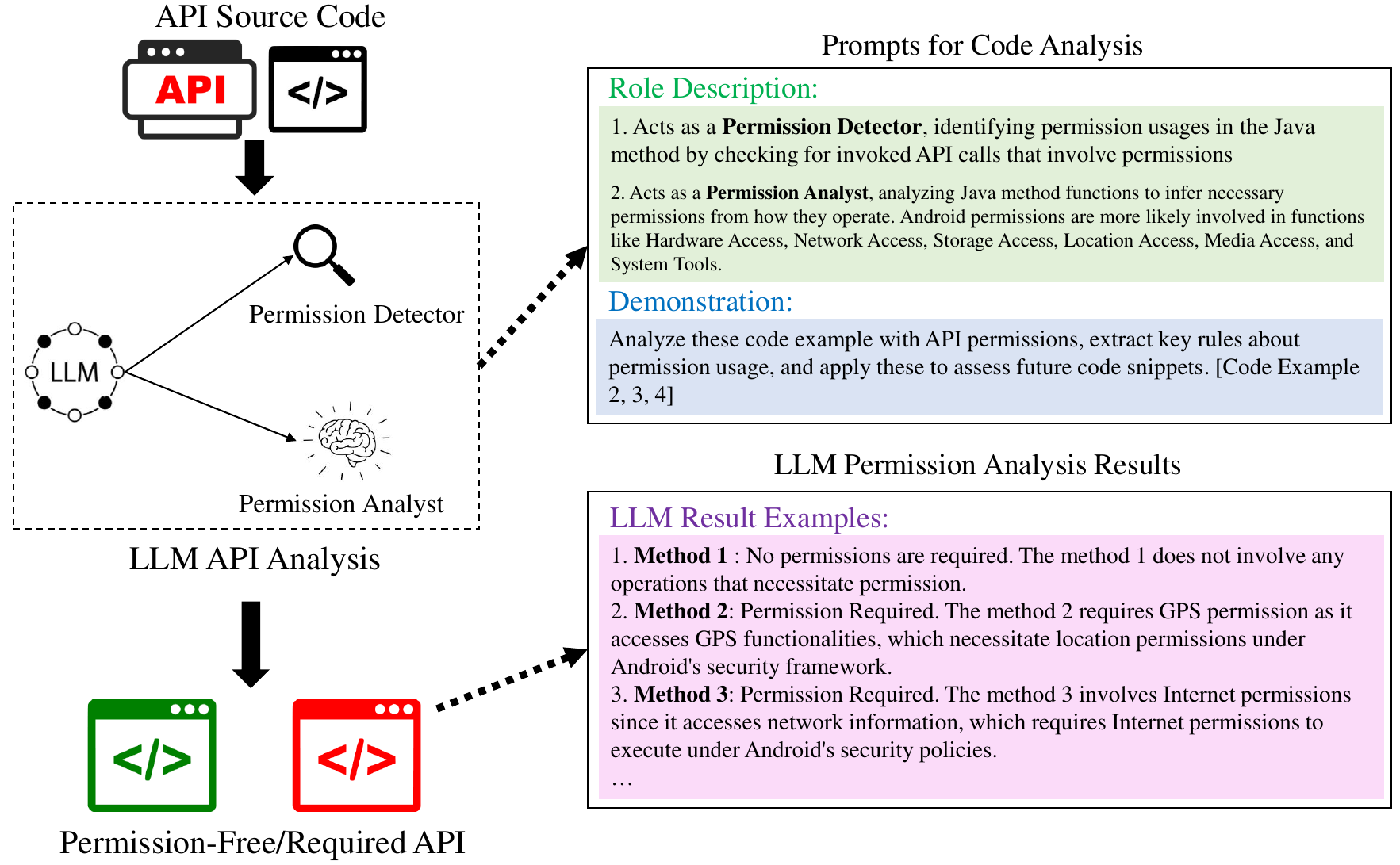}
\caption{Workflow of LLM API Analysis}
\label{fig:llmTemplate}
\end{figure}

In this phase, we utilize LLMs to analyze the Android SDK APIs extracted in the previous phase, and predict permissions required by each of the extracted APIs.
Figure~\ref{fig:llmTemplate} shows the workflow of LLM API analysis in which the dual-role prompting strategy employed, along with pre-demonstration cases for the LLMs to predict necessary permissions based on method signatures, method code and contextual documentation.

\subsubsection{Dual-Role Prompting Strategy}

The first aspect of our LLM-based framework for analyzing  API-permission mappings is the dual-role prompting strategy which configures the LLM to function in two distinct roles: as a code-based permission detector, and a code-based permission analyst.

\begin{figure}[htbp]
\centering
\includegraphics[width=\linewidth]{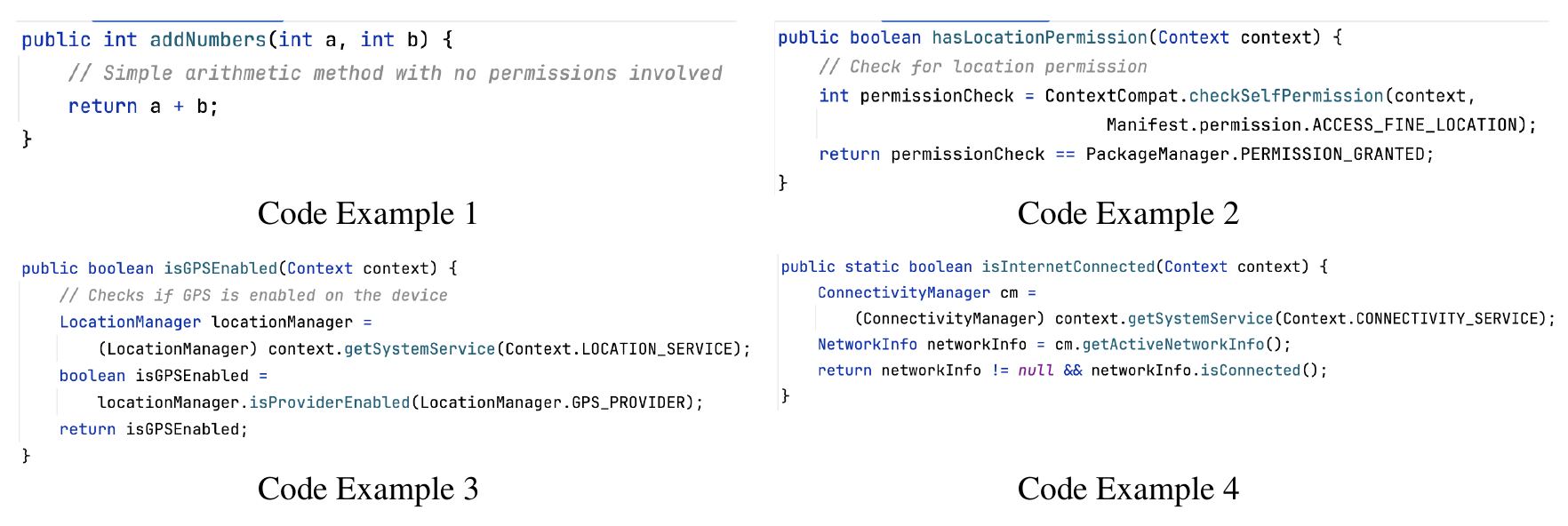}
\caption{Examples of Permission-Required and Permission-Free API}
\label{fig:codeExamples}
\end{figure}

Traditional API permission detection techniques have exhibited high efficacy for certain explicit APIs that expose their permission dependencies through semantic indicators in their names or accompanying comments. 
To continue leveraging this characteristic in our approach as well, we crafted a specialized role, code-based permission detector, for the LLM to ascertain the involvement of permissions by analyzing semantic information in the body of the method.
For instance, the Code Example 2, ``hasLocationPermission'', in Figure~\ref{fig:codeExamples} literally indicates a requirement for location permissions through its method name.

However, the effectiveness of these common detection methods is generally limited due to sparse documentation by Android on both the offical website or within the Android SDK source code. 
As such, we cannot expect all APIs to be accompanied by comprehensively descriptive comments or to follow a standardized naming convention. 
Figure~\ref{fig:codeExamples} presents Code Example 3 (\texttt{isGPSEnabled}) and Code Example 4 (\texttt{isInternetConnected}), 
which serve as illustrative instances within this category. These examples notably lack explicit mentions of permissions in comments and code, despite the necessity of specific permissions for accessing GPS and Internet functionalities in Android. 
Both Code Example 3 and Example 4 are permission-required APIs that do not possess clear semantics in their documentation or naming conventions.
Thus, the code-based permission analyst role of our dual-role prompting strategy is specifically catered to solving this gap. 
This role closely examines the body of API methods to understand their functionalities. According to the Android documentation~\cite{android_permissions_overview}, necessary permissions are typically associated with specific functionalities, including Hardware Access, Network Access, Storage Access, Location Access, Media Access, and System Tools. 

Our LLM prompts involve defining explicit role profiles and instructional prompts for each role. 
Figure~\ref{fig:llmTemplate} illustrates the template prompts we employ.
For the permission detector role, the prompt specifies: ``Acts as a Permission Detector, identifying permission usages in the Java method by checking for invoked API calls that involve permissions.''
Conversely, for the permission analyst role, the prompt directs the LLM to ``Act as a Permission Analyst, analyzing Java method functions to infer necessary permissions based on their operational characteristics.
Android permissions are more likely involved in functions like Hardware Access, Network Access, Storage Access, Location Access, Media Access, and System Tools.'' 

\subsubsection{Pre-demonstration Cases of LLMs}

Existing works that investigates In-Context Learning~\cite{rubin2021learning, tang2024context} has shown that high-quality pre-demonstration cases can significantly enhance the ability of LLMs to analyze code snippets~\cite{dong2022survey, zhang2023makes, gao2023makes}. 
By interacting with pre-annotated API-permission mapping examples, LLMs learns to recognize the patterns for invoking permission-requiring APIs, so as to develop an initial understanding of how permissions are implemented and invoked within the Android Framework. 
This understanding helps LLMs to identify implicit indicators of permission usage and infer correlations between invoked APIs and the requested permissions. 
Such capabilities are particularly useful in scenarios where API documentation lacks clear permission details.

Figure~\ref{fig:codeExamples} presents four strategically selected code examples that illustrate various scenarios involving API permission requirements or not.
Adapting the prompting template outlined in Figure~\ref{fig:llmTemplate}, these examples serve as demonstration prompts for the LLM. 
Each example includes a code snippet, accompanied by an optional code comment that describes the respective API functions. 
By reading the code snippets together with the accompanying annotations, the LLM better understands the relationship between the API calls and their corresponding permission requirements.
This template can be dynamically extended by modifying the example code presented to the LLM. 
Specifically, the extension process involves adding new API invocation code examples or adapting existing ones to represent a broader range of API usage scenarios for LLM to acquire a more generalized understanding.


\subsection{API Permission Verification}
After collecting predicted permission-required APIs within the Android SDK, we then automate the generation of self-contained API test cases to verify the predicted APIs.

\begin{figure}[htbp]
\centering
\includegraphics[width=0.9\linewidth]{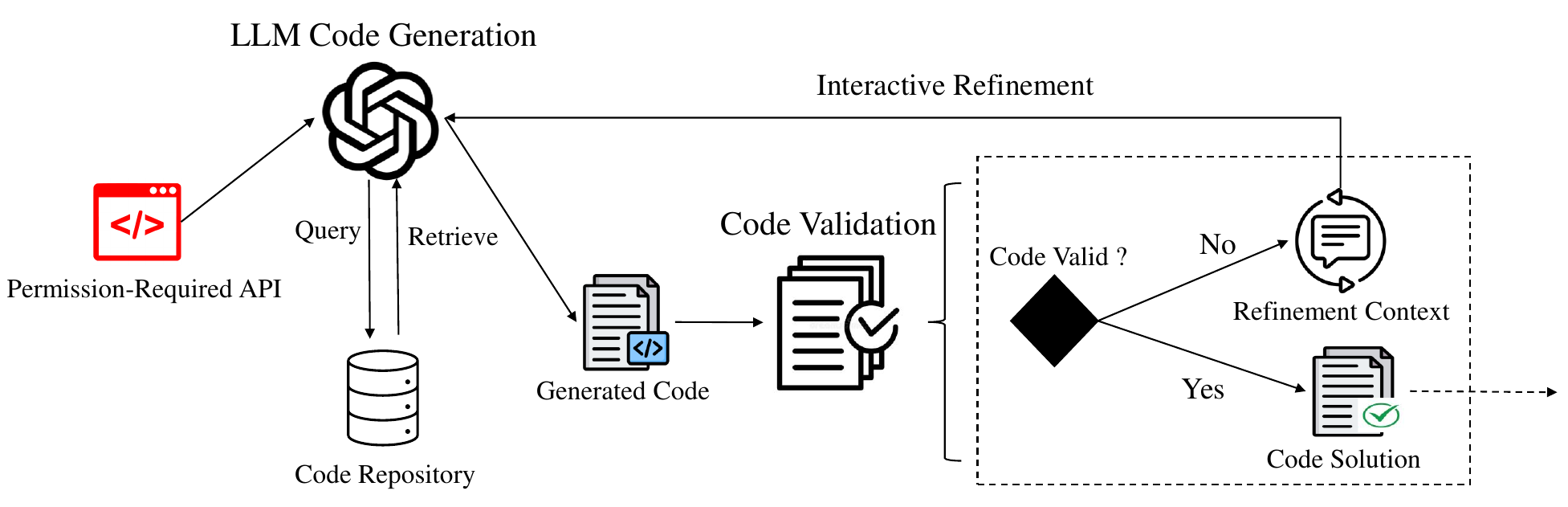}
\caption{API-Driven LLM Code Generation
}
\label{fig:codeGen}
\end{figure}

\subsubsection{API-Driven LLM Test Case Generation}
Our pipeline for API-driven LLM test case generation is illustrated in Figure~\ref{fig:codeGen}. To avoid duplicative efforts while ensuring the utility of our test cases, we adopt the Retrieval-Augmented Generation (RAG)~\cite{lewis2020retrieval} architecture to access existing code repositories, including BigCodeBench~\cite{zhuo2024bigcodebench} and Complexcodeeval~\cite{feng2024complexcodeeval}. Initially, we query the repository to determine whether there exists a pre-formed test case for the permission-required API. If an existing test case is found, it is normalized by the LLM into a self-contained format and returned. 
If no such test case exists, the LLM is prompted to generate a bespoke test case for the API.

\subsubsection{Verification by Code Validation Agent}
The Code Validation Agent, facilitated by an LLM, ensures that test cases comply with the self-containment rule and align with the SDK version being tested.
If a test case fails to meet these requirements, an iterative refinement process is initiated, where the Validation Agent provides feedback and additional contextual information to the LLM Code Generator.
This is repeated until the test case adheres to all specified criteria where the test case is considered to be finalized and delivered as a validated solution. 
During this process, due to the inherent variability and instability of current LLM outputs, manual intervention is occasionally required to refine and construct test cases that align with our specific requirements.


\subsubsection{Executing and Validating Test Cases on Emulator}
The final step of our methodology involves executing the generated test cases within a client app running in an emulated Android device. 
As specified in the Android SDK documentation~\cite{android_permissions_overview}, invoking an API without having the necessary permissions triggers a security exception for the client application. 
Therefore, we verify the accuracy of the API-permission mapping by observing whether the triggered exception message contains required permissions at the execution of the test case.

\section{Evaluation}
\label{sec:evaluation}

In this section, we address the three research questions (RQs) designed to assess the efficacy of our tool \tool{} as follows:



\textbf{RQ1: How effective is \tool{} compared to existing works?}
\begin{itemize}
    \item How effective is our tool \tool{} in identifying API-permission mappings within the Android Framework, in terms of the number of mappings discovered and its performance compared to existing static and dynamic analysis-based baselines?
\end{itemize}

\textbf{RQ2: How well does \tool{} perform when evaluated against Android SDK Source Code Annotation and Documentation?}
\begin{itemize}
    \item How effectively does our tool \tool{} identify API-permission mappings compared to the permission annotations found in Android SDK source code and official Android developer documentation? 
    
    \item To what extent do the SDK source code and official documentation omit necessary permission annotations and comments or include non-standardized annotations?
\end{itemize}

\textbf{RQ3: What insights can \tool{} provide about API-permission mappings across major Android Framework releases?}
\begin{itemize}
    \item What are the predominant characteristics and statistical patterns of API-permission mappings identified by \tool{} across different Android SDK versions?
    \item How do permission-required APIs evolve across successive Android Framework versions, and is there a specific example illustrating this progression?
\end{itemize}


\subsection{RQ1: How effective is \tool{} compared to existing works?}
For this research question, we select as comparison two state-of-the-art baselines: Dynamo~\cite{dynamo} and Arcade~\cite{arcade}. 
Both of the baseline approaches have published mappings for Android 6 which allows us to directly compare \tool{} to the two baseline. 
Arcade and Dynamo also report the number of covered APIs and discovered mappings for Android 7 and Android 10, respectively. Therefore, we compare the performance of \tool{} with Arcade for Android 7 and with Dynamo for Android 10.
Additionally, we further perform the API-permission mappings analysis on the latest stable Android 15, and publish the mappings.
Additionally, we extend our comparison to include NatiDroid~\cite{natidroid}, which is a tool specializing in mapping pairs between Java code and native C++ code within the SDK and have published API-permission mappings for Android 10. 


\subsubsection{Baselines}
Dynamo~\cite{dynamo} is the current state-of-the-art permission-required API detection tool, which revisits the imprecision issue in existing analysis approaches for Android API permission mapping, and delivers an improvement on existing works. 
Correspondingly, Arcade~\cite{arcade} has achieved the best performance via static analysis-based approach.
Arcade~\cite{arcade} adds path-sensitivity to static analysis for further precision improvement based on a novel graph abstraction technique.
Arcade extracts CFG representation of Android Framework, and derives a novel Access-Control Flow Graph which is processed to produce a succint representation of the access control conditions enforced by the API in the form of first-order logic.
Natidroid~\cite{natidroid} conducts an analysis of APIs spanning both Java and native C++, specifically addressing the permission aspects of cross-language APIs. 
In this context, we incorporate Natidroid~\cite{natidroid} to observe our approach's capability of analyzing permission requirements within cross-language interactions, notably between the Framework API in Java and the permission checks executed in native code.

\subsubsection{Experimental Settings}
Experiments are conducted on an emulated Pixel 5 device that comes with Android Studio, and is configured with 4 GB RAM.
The setup includes Android SDK versions 6, 7, 10, and 15  with an x86\_64 architecture system image. 
We employ ChatGPT-4 with the \texttt{gpt-4o-mini-2024-07-18} model as the LLM component of our methodology due to its advanced natural language processing capabilities, that is important for understanding code demostration examples for accurate API-permission mapping. 
The ChatGPT-4 API requires an average cost of approximately 50 USD and 25 hours to analyze all extracted Java methods within the SDK for a single version.
Additionally, we reuse experimental data previously published in existing works available on public websites for baseline comparisons wherever applicable.

\subsubsection{Experimental Results}

\begin{table}[ht]
\centering
\caption{Comparative Analysis of API-Permission Mappings across Different Tools. Arcade reports the number of covered APIs and discovered mappings for Android 7 in the paper, while Dynamo provides mappings exclusively for Android 10. As a result, we compare the performance of \tool{} with Arcade for Android 7 and with Dynamo for Android 10.}
\label{tab:api_permission_mappings}
\begin{tabular}{@{}llll@{}}
\toprule
Tool                  & Android SDK   & Covered API & Permission-Required API          \\ \midrule
Dynamo                & Android 6     & 2,057       & 1,294                  \\
Arcade                & Android 6     & 4,189       & 1,198 \\
\tool{}      & Android 6     & 9,406       & \textbf{2,234}                  \\  \midrule
Arcade   & Android 7     & 5,073      & 1,776                  \\
\tool{} & Android 7     & 11,875       & \textbf{3,552}                 \\
 \midrule
Dynamo   & Android 10    & 3,579       & 2,537                  \\
\tool{}      & Android 10    & 15,397      & \textbf{4,576}                  \\  \midrule
\tool{}      & Android 15    & 15,138      & \textbf{3,264}                  \\ \bottomrule
\end{tabular}
\end{table}

Table~\ref{tab:api_permission_mappings} presents the comparison between our tool, \tool{}, against established benchmarks Dynamo and Arcade in terms of the number of  covered APIs and discovered permission-requiring APIs across various versions of the Android Framework.

In Android 6, \tool{} identifies permission-required APIs in 2,234 out of 9,406 covered APIs. This represents a significant improvement over Dynamo, which identifies 1,294 permission-required APIs among 2,057 covered APIs, and Arcade, which finds 1,198 permission-required APIs across 4,189 APIs. These results shows that \tool{} achieves a broader coverage and detects a higher number of permission-required APIs compared to both state-of-the-art baselines. 

This improvement extends well for subsequent Android Framework version 7 and 10 where Arcade identifies 1,776 permission-required APIs in 5,073 covered APIs for Android 7, and our approach discovers 3,552 API-permission mappings among 11,875 covered APIs in Android 7. For Android 10, \tool{} uncovers 4,576 permission-required APIs out of 15,397 covered APIs while Dynamo covers 3,579 APIs and identifies 2,537 permission-required APIs in the same framework version. 
The improvement in the detection of permission-requiring APIs compared to the baselines is shown across multiple Android Framework versions underscoring \tool{}'s scalability and adaptability. 
\tool{} identifies 3,264 permission-required APIs out of 15,138 covered APIs in Android 15, indicating a decreasing trend in both the total number of covered APIs and those requiring permissions compared to previous versions. 
This trend will be further investigated in RQ3 to elucidate potential underlying causes.


\subsubsection{Overlap and Novelty in API Discoveries}
\begin{table}[ht]
\centering
\caption{Detailed Comparison of API-Permission Mappings Across Tools}
\label{tab:api_mappings_detail}
\begin{tabular}{@{}lllll@{}}
\toprule
Tool            & Android SDK & Total & Same API & New API \\ \midrule
Arcade          & Android 6   & 1,198 & 929    & 1,305     \\
Natidroid         & Android 10  & 282   & 264      & 34      \\ \bottomrule
\end{tabular}
\begin{tablenotes}
\item \textbf{Note:} ``Same API'' refers to APIs identified by both the compared baselines and \tool{}. ``New API'' refers to APIs identified exclusively by \tool{} that are not detected by the compared baselines.
\end{tablenotes}
\end{table}

Arcade publicly disclosed all API-Permission mappings identified in Android 6, allowing us to directly compare our findings with theirs. Similarly, Natidroid released the cross-language API-Permission mappings discovered in Android 10, enabling a comparative analysis with our results.
Table \ref{tab:api_mappings_detail} provides a detailed comparison of API-permission mappings as identified by \tool{} relative to Arcade and Natidroid. 
This comparison not only considers the total APIs identified by each tool but also examines the overlap, and new discoveries unique to each method.

\textbf{Comparison with Arcade (Android 6):}
According to Table~\ref{tab:api_permission_mappings}, \tool{} identifies a total of 2,234 permission-requiring APIs, compared to Arcade, which identifies 1,198 permission-requiring APIs in Android 6. 
We can see from Table~\ref{tab:api_mappings_detail} that there is an overlap of 929 APIs exists between 2,234 APIs discovered by \tool{} and 1,198 APIs by Arcade.
Arcade detects 269 APIs that \tool{} does not capture. Conversely, \tool{} identifies 1,305 new APIs not previously captured by Arcade. 
The considerable overlap validates the reliability of \tool{}, and the discovery of new APIs highlights the capability of our tool \tool{} in uncovering more API-permission mappings.

\textbf{Comparison with Natidroid (Android 10):} 
The Android Native Development Kit (NDK) reference webpage documents the currently identified cross-language Android APIs~\cite{androidndk2025}. Within the Android SDK source code, Java methods located in the \texttt{frameworks/base/core/java/} sub-package and annotated with the \texttt{native} keyword constitute the Java Native Interface (JNI), which facilitates communication between Java and C++ code.
Consequently, we employ these documented characteristics as criteria to validate whether a method qualifies as part of the cross-language API mappings.
In the context of cross-language API mappings, Natidroid identifies a total of 282 APIs, of which 264 overlap with the results produced by \tool{}. This substantial overlap indicates that \tool{} effectively captures the majority of significant cross-language mappings identified by Natidroid.
Moreover, our analysis identifies 34 cross-language APIs in \tool{}'s results that Natidroid fails to detect, highlighting the enhanced detection capabilities of our approach in the complex domain of cross-language API-permission mappings between native C++ and Java APIs.

\rqbox{
\textbf{Answer to RQ1:} Our experiments demonstrates that \tool{} outperforms the existing state-of-the-art approaches in the number of mappings discovered both in terms of traditional same-language API-permission mappings, and also in cross-language context.
Its capability to uncover a substantial number of previously undetected permission-required APIs, when compared to baseline methods, shows that \tool{} provides deeper insights and wider coverage. 
This enhances the robustness of security configurations within the Android SDK environment. Compared to traditional dynamic and static analysis techniques, \tool{} offers greater flexibility and superior efficacy, consistently delivering the most comprehensive results in identifying API-permission mappings with a broadly accessible technology in emulator-based dynamic analysis.
}

\subsection{RQ2: How well does \tool{} perform when evaluated against Android SDK Source Code Annotation and Documentation?}


Pioneering studies in permission specification analysis~\cite{dynamo} have established the major inconsistency issues between source code comments and annotations, and the official documentation website. 
This RQ investigates how \tool{}'s API-permission mapping findings can be leveraged to improve these documentation practices.


\subsubsection{Documentation Practice in Source Code of Android SDK}
\label{sec:protocol}
Official Android documentation specifies that, starting with Android 6.0 (API level 23), Google has formalized the documentation of permission specifications through two principal methods~\cite{android_permissions_overview}:
\begin{enumerate}
    \item The use of the Java annotation \texttt{@requiresPermission} to associate APIs with specific permissions.
    \item The application of the \texttt{@link android.Manifest.permission\#} annotation to explicitly detail the permissions required by an API.
\end{enumerate}
Following the above protocol, our study parses the entire source code of the Android SDK~\footnote{For clarification reasons, SDK is different from Android Framework; Android Framework is a middleware that communicates with Android applications inside the Android operating system, whereas Android SDK provides stubs/APIs to communicate with the Android Framework during the development and compilation of Android applications.} to extract permission annotations from all method definitions. 
Specifically, we focus on identifying and analyzing APIs that incorporate the \texttt{@requiresPermission} annotation. Those APIs lacking this annotation are categorized under missing permission declarations, and will be documented as evidence for incomplete API-permission mappings of the Android Framework documentation.

\subsubsection{Web Documentation Practice of Android SDK}

The official documentation website for Android SDK~\cite{android_reference} is found to be lacking in a standardized method for documenting permissions required by APIs.
For example, some pages and sections uses the \texttt{@requiresPermission} annotation similar to the annotation found inside the source code, others state the permission only inside the text description of the API's section.
The latter can be deemed as a violation of documentation protocol described in \ref{sec:protocol} for not dedicating a subsection to annotate \texttt{@requiresPermission} to provide a predictable documentation format that is conducive to web scraping by Android app developers and security analysts.
APIs without these annotations are classified as non-standardized permission-required APIs, distinguishing them from those with missing permissions in the source code.

\subsubsection{Results Analysis}

\begin{table}[ht]
\centering
\caption{Comparison of API-Permission Mapping Discoveries Across SDK Versions}
\label{tab:vsDoc}
\scriptsize 
\begin{tabularx}{\linewidth}{@{}lXXXXXX@{}} 
\toprule
SDK Version & \multicolumn{1}{c}{\tool{}} & \multicolumn{2}{c}{SDK Source Code Annotation} & \multicolumn{2}{c}{Official Web Documents} \\
\cmidrule(lr){2-2} \cmidrule(lr){3-4} \cmidrule(lr){5-6}
& Discovered APIs & Annotated APIs & New Disc. (\tool{}) & Annotated APIs & New Disc. (\tool{}) \\
\midrule
Android 7 & 3,552 & 65 & 3,487 & 13 & 3,539 \\
Android 10 & 4,576 & 698 & 3,906 & 57 & 4,519 \\
Android 15 & 3,265 & 1,076 & 2,202 & 165 & 3,100 \\
\bottomrule
\end{tabularx}
\begin{tablenotes}
\item \textbf{Note:} The \textit{New Disc. (LLM)} column does not represent the difference between \textit{Discovered APIs} and \textit{Annotated APIs}. Instead, it indicates the number of new API-permission mappings uniquely identified by \tool{}, which are not annotated in source code or official web documentation.
\end{tablenotes}
\end{table}

We then conduct a detailed evaluation of the outcomes from our study, comparing our \tool{} against SDK Source Code Annotation and Official Web Documentation in the discovery of API-permission mappings across different Android SDK versions.
Table~\ref{tab:vsDoc} presents these findings. 
The column labeled \textit{Discovered APIs} presents the number of permission-requiring APIs uncovered through our \tool{}. Meanwhile, the columns titled \textit{Annotated APIs} denote the count of permission-annotated methods identified in the Android SDK source code and those documented in the official online Android resources respectively. 
The \textit{New Disc. (\tool{})} column does not represent the difference between \textit{Discovered APIs} and \textit{Annotated APIs}. Instead, it indicates the number of new API-permission mappings identified by \tool{}, which are not annotated in source code or in official web documentation respectively.

 The scope of our experiment contains 3 Android SDK versions, and the results of our experiments reveal that the number of API-permission mappings identified across the three Android SDK versions varies significantly:
\begin{itemize}
    \item \textbf{Android 7:} \tool{} discovered 3,552 APIs, significantly exceeding the 65 annotated in the source code and the 13 documented in the official web documentation. We discover 3,487 additional mappings compared to existing source annotations and 3,539 beyond official web documentation.
    \item \textbf{Android 10:} \tool{} identified 4,576 APIs, compared to 698 annotated in the source code and 57 standardized in the web documentation, uncovering 3,906 additional mappings beyond source code annotations and 4,519 beyond web documentation.
    \item \textbf{Android 15:} \tool{} discovered 3,265 APIs, compared to 1,076 annotated in the source code and 165 documented in the web documentation, demonstrating its ability to reveal 2,202 additional mappings over source code annotations and 3,099 over official documentation.

\end{itemize}


\rqbox{
\textbf{Answer to RQ2:} \tool{} uncovers clear deficiencies in current API documentation practices, which struggle to keep pace with rapid technological advancements, and suggests a path toward more structured and reliable strategies.
The juxtaposition of \tool{}'s mappings with those given by annotations inside the SDK source code and the documentation on Android SDK official website further establishes the advantage in effectiveness of LLM-driven methodologies over traditional approaches in the discipline of API permission specification. 
This advantage enabled by cutting edge technologies like LLM will profoundly impact the effectiveness and efficiency of downstream tasks in software and security analysis that relies on the published API-permission mappings.
Promising findings in this paper also advocates for further exploration into integrating LLM with other traditional software development and documentation processes.
}

\subsection{RQ3: What insights can \tool{} provide about API-permission mappings across major Android Framework releases?}


This RQ investigates permission specification in a different dimension, across different Android SDK versions: 7, 10, and the latest stable 15.
Android 7 and 10 are the version that prior studies performed their evaluations on and published API-permission mappings for.
Android 15 is the latest version of Android that is the most relevant for studying the most recent API management practices.
Our study examines the distribution of API-permission mappings within various packages across these versions, showing the evolutionary trends and identifying key areas of interest for both developers and security analysts.

\begin{table}[ht]
\centering
\caption{Distribution of API-Permission Mappings Across Android Versions}
\label{tab:api_distribution}
\begin{tabular}{@{}lccc@{}}
\toprule
\textbf{Package} & \textbf{Android 7} & \textbf{Android 10} & \textbf{Android 15} \\ \midrule
android          & 1905              & 2075                & 1631               \\
com              & 1391              & 2097                & 1116               \\
java             & 212               & 360                 & 328                \\
org              & 3                 & 6                   & 56                 \\
javax            & 37                & 25                  & 24                 \\
sun              & -                 & -                   & 67                 \\
jdk              & -                 & 6                   & 7                  \\
libcore          & -                 & -                   & 12                 \\
gov              & -                 & -                   & 24                 \\
jsr166           & 4                 & 4                   & -                  \\
androidx         & -                 & 3                   & -                  \\
\midrule
\textbf{Total Mappings}   & 3552              & 4576                & 3265               \\
\bottomrule
\end{tabular}
\end{table}

\subsubsection{Quantitative Analysis}
As depicted in Table~\ref{tab:api_distribution}, our findings 3,552, 4,576, and 3,265 permission-requiring APIs in the SDKs of Android 7, 10 and 15 respectively.
Table~\ref{tab:api_distribution} also shows the distribution of discovered API-permission mappings in three Android versions across all identified packages:
\begin{itemize}
    \item \textbf{Android 7:} Dominated by the \texttt{android} and \texttt{com} packages with 1,905 and 1,391 mappings respectively, indicating the focus of permission-driven protection for APIs that communicate with Android-related packages. 
    \item \textbf{Android 10:} Shows a notable increase in permissions within the \texttt{com} package (2,097 mappings) within which a lot of third-party packages also resides. Android 10 also saw with the introduction of the \texttt{androidx} package, and consequently, permission-driven protections for its APIs.
    \item \textbf{Android 15:} Highlights a more balanced distribution across packages, with significant permissions mapped within \texttt{android} (1,631 mappings) and \texttt{com} (1,116 mappings), alongside the emergence of permissions in the \texttt{sun} and \texttt{gov} packages indicating a shift in development practices towards newer libraries and frameworks.
\end{itemize}

We learn from the distribution of across the packages and Android versions that \texttt{android}, \texttt{com}, and \texttt{java} are the three principal packages containing APIs that perform the most security-sensitive operations within the Android ecosystem.
The  \texttt{android} package encompasses the core of the Android platform, packaging core functionalities for Android app development such as managing user interface components, application lifecycle management, and system services. 
Moreover, it also contains the procedures for essential device functions such as cameras, sensors, and storage, communications, security, and permissions. 
Given central role in the Android architecture played by procedures its subpackages, it makes sense that the \texttt{android} package contains the most permission-requiring APIs compared to other packages as those procedures encapsulate operations that are security-sensitive and should not be accessible by users and processes that are not granted necessary permissions.

Conversely, the \texttt{com} package predominantly comprises classes from third-party libraries, including those by Android vendors such as Samsung, and Vivo.
These packages look to extend the functionalities that come with the standard vanilla Android Framework, by incorporating their own APIs and procedures.
For instance, Google services such as Maps do not come automatically with Android Open Source Project, and are organized under \textit{com.google.} subpackages, and Android features that integrate with Google services \texttt{com.google.android} are also found under \texttt{com.google.} subpackage.

Lastly, the \texttt{java} package provides foundational yet important classes that are utilized across various Java-based environments.
Although it is less directly engaged with device-specific functionalities compared to the \texttt{android} package, the APIs within the \texttt{java} package still provides functionalities for low-level operations such as networking and I/O that necessitate permission-driven protection to safeguard sensitive resources in the device. 

\begin{figure}[htbp]
\centering
\begin{tikzpicture}
\begin{axis}[
    ybar stacked,
    bar width=15pt,
    enlarge x limits=0.25,
    ylabel={Number of Discovered Mappings},
    symbolic x coords={Android 7, Android 10, Android 15},
    xtick=data,
    x tick label style={rotate=45, anchor=east},
]
\addplot+[ybar] plot coordinates {(Android 7,1905) (Android 10,2075) (Android 15,1631)};
\addplot+[ybar] plot coordinates {(Android 7,1391) (Android 10,2097) (Android 15,1116)};
\addplot+[ybar] plot coordinates {(Android 7,212) (Android 10,360) (Android 15,328)};
\legend{android, com, java}
\end{axis}
\end{tikzpicture}
\caption{API-permission mappings by major package across different SDK versions.}
\label{fig:apiDistribution}
\end{figure}
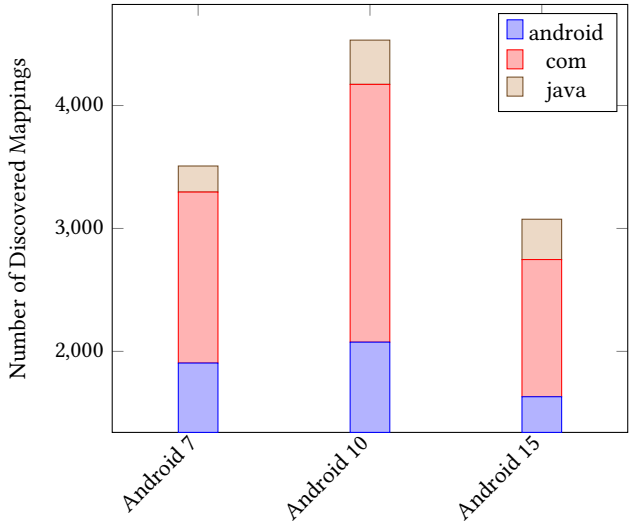

As we seek further insights into the evolution of package-permission relationship in Android Framework, we visualize in Figure~\ref{tab:api_distribution} the distribution of API-permission mappings within the three core packages across Android SDK versions 7, 10, and 15. 
It highlights a decrease in the number of APIs requiring permissions in the \texttt{android} and \texttt{com} packages, while \texttt{java} package remains relatively stable. 
This can be considered an intriguing shift within the Android ecosystem in terms of development and security practice that is worth investigating.

\textbf{Android Package (\texttt{android}):}
We have deduced several following factors that explains the observed reduction in the number of API-permission mappings in Android 15:
\begin{itemize}
  \item \textit{API Optimization and Consolidation:} Workflow optimization for communication with the Android Framework leads to consolidation of existing APIs to reduce permission bloat across the APIs.
  \item \textit{Increased Security Measures:} More rigorous security protocols in newer Android versions limits third-party packages' access to sensitive procedures, and thus reducing the number of permission-requiring APIs in those packages.
  \item \textit{Deprecation of Older APIs:} The periodic deprecation of APIs in favor of newer, more secure, and efficient alternatives contributes to fewer permission-requiring APIs from older Android versions.
\end{itemize}

\textbf{Com Package (\texttt{com}):}
A sharp reduction in the number of permission-requiring APIs within the \texttt{com} package indicates significant adjustments in the integration with third-party library:
\begin{itemize}
  \item \textit{Removal of Redundant or Unsafe APIs:} To enhance security and development efficiency, third-party libraries may phase out APIs that are either obsolete or pose security risks.
  \item \textit{Optimization of Library Code:} Just like with the packages inside \texttt{android}, optimization of third-party library code could include consolidation of procedures and reassignment of permissions that result in fewer permissions being required.
\end{itemize}

\textbf{Java Package (\texttt{java}):}
The stability with regards to the number of API-permission mappings in the \texttt{java} package across the examined SDK versions underscores the core part it plays to provide offering essential functionalities in Android:
\begin{itemize}
  \item \textit{Mature API Set:} The APIs under the \texttt{java} package are well-established compared to those inside \texttt{android} and third-party libraries inside \texttt{com} and thus, have stable permission requirements, which leads to less change in the number of mappings across Android versions.
  \item \textit{Less Interaction with System Features:} Java APIs generally do not access permission-requiring critical system functionalities as it sits at a lower level than \texttt{android} and \texttt{com} libraries.
\end{itemize}


\subsubsection{Qualitative Analysis}
We perform a qualitative analysis by examining the source code of specific packages in the Android SDK.
\paragraph{\textbf{Server Package}}
We noticed fluctuations in the number of API-permission mappings within the \texttt{/com/android/server} package.
\tool{} identified 567, 1029, 956, and 96 mappings across Android SDK versions 7, 10, 14, and 15 respectively. 
This sharp decline in mappings from Android 14 and Android 15 warrants a closer inspection of the source code where we found that this change is primarily due to extensive refactoring of procedures by the developers.
For instance, the refactoring substantially altered the design of web server-related APIs and underlying procedures within the Android Framework through the consolidation of functions to diminish the need for redundant permission requests across nested method calls, especially involving Internet connection.
This establishes the Android community's proactiveness in undertaking architectural revisions that allows Android to maintain balance in functionality and security.

\paragraph{\textbf{Sun, gov and libcore Packages}}
In Android 15, we observe a clear shift in the distribution of API-permission mappings across packages, namely the introduction of mappings in \texttt{sun}, \texttt{gov} and \texttt{libcore}, and the reduction of mappings in other packages. 
For instance, API-permission mappings started to get discovered in the \texttt{sun} package which is traditionally associated with low-level system operations. 
This change reflects possible system integrations or enhancements in security features necessitating more explicit permissions. 
Similarly, the \texttt{gov} package, which is ostensibly government-specific applications, also starts to show mappings in Android 15.
This indicates an increased focus on mobile solutions for government services that require heightened security protocols and access controls. 
Furthermore, the \texttt{libcore} package, which provides core libraries for the Java programming language, and the \texttt{jdk} (Java Development Kit) package, essential for Java applications, both have an increase in mappings, indicating broader utility or security updates that demand additional permissions.

\paragraph{\textbf{Jsr166 and androidx Packages}}
Conversely, we observe a decrease in the number of API-permission mappings in the \texttt{jsr166} and \texttt{androidx} packages. 
The \texttt{jsr166} package provides concurrency utilities that might have undergone enhancements that reduce the necessity for explicit permission checks, as part of its maintenance for improving efficiency and, at the same time, maintaining security.
The \texttt{androidx} package replaces the original Android support libraries could have undergone API deprecations to minimize bloat and improve adherence to software development best practices, and subsequently, end up removing existing permission requirements.


We observe a pattern which includes an increase in API-permission mappings in Android 10, followed by a reduction in Android 15. The latter could indicate a phase of consolidation or optimization in the Android Framework which could be caused by the reevaluation of permission requirements as part of the a security On the other hand, the emergence of permissions in specialized packages such as \texttt{sun} and \texttt{gov} in Android 15 suggests a more diverse API usage scenarios that is enabled by the introduction of new features by third-party developers or regulatory mandates by governing authorities respectively.

\rqbox{
\textbf{Answer to RQ3:} This analysis of the mappings' evolution across Android version sheds light on the dynamic security landscape of the Android ecosystem. 
More insights are also uncovered through the close-up analysis of the distribution of mappings, as it indicates changes in development practices and dependency management that are critical for secure and performant Android apps. 
Understanding trends in both of these dimensions allow Android app and platform developers to adhere to secure and efficient software development practices, thereby mitigating vulnerabilities and ensuring security of users and their devices.
}

\section{Threat to Validity}
\label{sec:discussion}

\paragraph{\textbf{Impact of API Coverage on Permission Detection}}
One potential threat to internal validity in this study arises from the difference in the number of APIs covered between \tool{} and the baseline approaches. 
In RQ1, \tool{} performed permission prediction and validation over a significantly higher number of candidate APIs compared to other baseline tools.
\tool{} achieves its higher coverage by a combination of static AST parsers and LLM that enables \tool{} to analyze more APIs in SDK, and tracking both documented and undocumented APIs that may be overlooked by traditional permission mapping analysis techniques.
This broader coverage gives \tool{} a bigger pool of APIs that may contain a greater number of permission-required APIs in terms of absolute numbers; however, it may also introduce an bias in measuring the extent of LLM's effectiveness due to the coverage intrinsically being tied to the results.

Nevertheless, it is important to note that the primary objective of this study is to uncover as many valid API-permission mappings as possible within the Android SDK. 
From this perspective, the core goal of the comparison is not to evaluate the tools under constrained API coverage but to determine which tool as a whole is more effective in identifying more valid mappings inside the Android Framework at the end of the day. 
This aligns with the purpose of \tool{} as an complete package for the extraction of comprehensive API-permission mapping rather than narrowly evaluating the LLM component of the methodology.

\paragraph{\textbf{Impact of Inherent Instability of LLM Outputs}}
The inherent instability of LLM outputs could also pose a threat to the internal validity of this study. 
To mitigate this, we implemented a dual-role and interactive refinement process during the API analysis and code generation stages of \tool{}. 
Both dual-role and interactive refinement enhances the stability and reliability of the results, while also preserving the robustness in model outputs to cover a diverse set of test cases.


\paragraph{\textbf{Impact of Human Intervention on Code Generatio}n} 
In some API cases, manual human intervention is required during the code generation process. 
This is due to the limitations of LLMs in understanding complex code structures, despite recent advancements. 
As a result, code generation models are improving to reduce human assistance, but completely removing the need for intervention is still challenging.

\section{Conclusion and Future Work}
\label{sec:conclusion}
This paper introduced a novel for API-permission mapping of the Android Framework, and proposed \tool{}, an LLM-based tool, that performs static analysis on the Android SDK, and dynamic analysis using API code generation. 
We formulated three research questions aimed at evaluating the performance of \tool{} relative to existing state-of-the-art baselines, assessing the quality of official Android documentation, and analyzing the trends and characteristics of API-permission mappings across various SDK versions.
Our experiment results show that our tool \tool{} surpasses existing static and dynamic analysis baselines in effectiveness for identifying API-permission mappings inside the Android Framework. 
We also identified shortcomings in the official Android documentation in terms of the completeness of API-permission mappings provided to the Android application developers. 
Finally, we observed that the Android Framework has undergone substantial evolutions across major releases as shown by the fluctuation in the number API-permission mappings in various packages inside the Android Framework.
Future research potential lies in improving the robustness of LLM models' outputs and utilize \tool{} to maintain API-permission mappings for more versions of Android SDKs.

\bibliographystyle{ACM-Reference-Format}
\bibliography{references}

\appendix

\end{document}